\documentclass[aps,prl,twocolumn,amsmath,amssymb]{revtex4}
\usepackage{graphicx}
\usepackage{amsmath}
\usepackage{amsbsy}
\usepackage{dcolumn}
\usepackage{bm}

\begin{document}
\noindent
{\bf Comment on ``Fano Resonance for Anderson Impurity Systems''}

In a recent Letter, Luo {\it et al.} \cite{luo04} analyze the Fano line 
shapes obtained from scanning tunneling spectroscopy (STS) of transition
metal impurities on a simple metal surface, in particular of the
Ti/Au(111) and Ti/Ag(100) systems \cite{crommie00}. 
As the key point of their analysis, they claim that there is not only 
a Fano interference effect between the impurity d-orbital and the 
conduction electron continuum, as derived in Ref.\ \cite{ujsaghy00},  
but that the Kondo resonance in the d-electron spectral density 
has by itself a {\it second} Fano line shape, leading to the 
experimentally observed spectra. In the present note we
point out that this analysis is conceptually incorrect. Therefore,
the quantitative agreement of the fitted theoretical spectra with the
experimental results is meaningless. 
The nomenclature and the equation numbers below refer to Ref.\ \cite{luo04},
unless stated otherwise.

Luo {\it et al.} adopt an effective Fermi liquid (FL) picture, 
where the impurity spectrum is
comprised of three well-defined quasiparticle states, the 
single-particle levels at energies $\varepsilon _d$ and 
$\varepsilon_d +U$, respectively, and the Kondo resonance near the 
Fermi energy $\varepsilon _F$. 
The latter arises from resonant spin flip scattering of the
impurity electrons from the conduction electrons, induced by the
interplay of hybridization $V$ and on-site Coulomb interaction
$U$. Luo {\it et al.} describe this phenomenologically 
by means of a Dyson equation (4) which defines the 
potential scattering T-matrix 
$T_d(\omega)$ of the impurity electrons. We point out that a 
microscopic derivation of Eq.\ (4) does not exist, since 
interaction effects ($U$, $V$) are to some degree already incorporated
in Luo {\it et al.'s} ``bare'' impurity Green's
function $G_d^0(\omega)$, Eq.\ (5), and therefore Wick's theorem 
is not valid here. Nevertheless, such an effective treatment 
may be valid in the Kondo FL regime, where the high and low energy 
scales are separated. However, the implications drawn from it 
in \cite{luo04} are erroneous in three major points. 

(1) FL theory \cite{hewson93} as well as numerous exact NRG calculations 
(see, e.g., Ref.\ \cite{costi96}) demonstrate
that in the Kondo as well as in the mixed valence regime
the full impurity Green's function $G_d(\omega )$ has a simple 
peak structure near $\varepsilon _F$. It has, in general, a slight 
asymmetry caused by potential scattering, but by no means the
peak-dip shape of a Fano resonance. Therefore, 
$G_d(\omega )$ is approximately given near $\varepsilon _F$ by
\begin{eqnarray}
G_{d\sigma}(\omega) 
\approx \frac{\Gamma _K/\Delta}{\omega - \varepsilon _K + i\Gamma_K},
\quad |\omega| \lesssim T_K,\ T=0.
\end{eqnarray}
$T_d(\omega )$ is then uniquely determined by Eq.\ (4),
i.e. it has a complicated energy dependence which
can be extracted from Eq.\ (1) of the present Comment 
in combination with Eqs.\ (4), (5). In contrast, Luo {\it et al.}
assume {\it ad hoc}, c.f. Eq.\ (7), 
that $T_d(\omega )$ has the energy dependence
of Eq.\ (1) of this Comment, in clear contradiction to the correct
statements given above.
To illustrate this fact, we point out that Eq.\ (8) reduces to
the known Kondo resonance form only in the particle-hole symmetric case
($U=2|\varepsilon_d|$), or in the limit $\Delta/|\varepsilon_d| \ll 1$,
$U \gg 2 |\varepsilon _d|$, 
where $q_d(\omega)$ diverges at the position of the Kondo resonance,
but fails to do so for all other cases in the Kondo regime.
Consequently, the subsequent fitting of the experimental spectra is invalid.

(2) Luo {\it et al.} describe the 
Ti/Au(111) and Ti/Ag(100) systems \cite{crommie00} 
as being in the mixed valence regime, 
($\Delta/\varepsilon _d \approx 1$), where their Fano-like 
factor $q_d(\omega)=-{\rm Re}G_d^0(\omega)/{\rm Im}G_d^0(\omega)$
is finite near $\varepsilon _F$.
However, in this regime, there is no separation
between low-energy spin-flip scattering and high-energy Coulomb and
hybridization processes, which is the basis for Luo {\it et al.'s} 
analysis. Instead, in the mixed valence regime, the $\varepsilon_d$ peak 
of the impurity spectrum and the Kondo resonance are known 
(e.g. from exact NRG calculations \cite{hewson93,costi96}), 
to merge to a single
peak near $\varepsilon _F$ of width $\sim \Delta$, and there is no distinction 
between $\Delta$ and $\Gamma_K$, again in contrast to Eq.\ (8).
We note in passing, that the quantitative agreement of Luo {\it et
al.'s} results with NRG calculations shown in their Fig. 3
pertain only to quantities which are determined by high energy
physics and are, thus, not relevant for the present points of
criticism. 

(3) The fitting of Luo {\it et al.'s} expressions to the 
Ti/Au(111) and Ti/Ag(100) systems yields a width
of the ``mixed valence'' resonance peak of $\Delta$=54.5 meV 
and $\Delta$=29.2 meV, respectively. 
This is inconsistent with the expectation that for transition metal
impurities on metal surfaces the width $\Delta$ of the single-particle 
resonance is at least one order of magnitude larger.
The narrow widths of the 
experimental spectral peaks suggests, that these are, in fact, Kondo
peaks. We have theoretical evidence, that the complex line shape 
observed in Ti/Au(111) and Ti/Ag(100) \cite{crommie00}
is due to the multiple Kondo
resonances arising from the crystal-field split local orbitals of the
transition metal impurity. 
This will be discussed elsewhere. ---
This work was supported in part by DFG through SFB 608 and through
KR1726/1.

\vspace{0.2cm}
\noindent
Ch. Kolf$^1$, J.\ Kroha$^1$, M. Ternes$^2$, and W.-D.\ Schneider$^2$ \\
{\small
${}^1$Physikalisches Institut, \\
\hphantom{${}^1$}Universit\"at Bonn, 53115 Bonn, Germany\\ 
${}^2$Institut de Physique des Nanostructures,\\ 
\hphantom{${}^2$}Ecole Polytechnique F\'ed\'eral de Lausanne, \\
\hphantom{${}^2$}CH--1015 Lausanne, Switzerland\\
\noindent
Received 24 March, 2005}

\end{document}